\documentclass[]{raa}            
\usepackage{graphicx,times}
\usepackage{natbib}

\providecommand{\bm}[1]{\mbox{\boldmath$#1$\unboldmath}}

\newcommand{\1}{\Omega_\mathrm{M0}}
\newcommand{\2}{\Omega_\mathrm{r_{\rm c}}}

\begin{document}

   \title{Cosmological Constraints on the DGP braneworld model with Gamma-ray bursts
}

 \volnopage{ {\bf 2011} Vol.\ {\bf 11} No. {\bf 5}, 497--506}
   \setcounter{page}{1}

   \author{Nan Liang
        \and Zong-Hong Zhu
        }

   \institute{Department of Astronomy, Beijing Normal   University, Beijing 100875, China \\
   {\it liangn@bnu.edu.cn; zhuzh@bnu.edu.cn}\\
\vs \no
   {\small Received 2010 October 13; accepted 2011 January 4 }
}

\abstract{We investigate observational constraints on the Dvali,
Gabadadze and Porrati (DGP) model with Gamma-ray bursts (GRBs) at
high redshift obtained directly from the Union2 Type Ia supernovae
data (SNe Ia) set. With the cosmology-independent GRBs, the Union2
set, as well as the cosmic microwave background (CMB) observations
from the WMAP7 result, the baryon acoustic oscillation, the baryon
mass fraction in clusters and the observed $H(z)$ data, we obtain
that the best-fit values of the DGP model are $\{ \1,\2 \}
=\{0.235_{-0.014}^{+0.015},0.138_{-0.048}^{+0.051}\}$, which favor a
flat universe; and the transition redshift of the DGP model is
$z_\mathrm{T}=0.67_{-0.04}^{+0.03}$. These results lead to more
stringent constraints than the previous results for the DGP model.
  }

\authorrunning{Nan Liang and Zong-Hong Zhu}
\titlerunning{Constraints on the DGP model with Gamma-ray bursts}
   \maketitle


%

%

\section{Introduction}
The accelerating expansion of the current universe has been
confirmed by recent cosmological observations, such as Type Ia
supernovae (SNe Ia; Riess et al. 1998; Perlmutter et al. 1999;
Amanullah et al. 2010), cosmic microwave background (CMB; Bennett et
al. 2003; Spergel et al. 2003; Komatsu et al.  2010),  large scale
structures (LSS, Tegmark et al. 2004; Eisenstein et al. 2005), as
well as the x-ray gas mass fraction of clusters (Allen et al. 2004).
By assuming General Relativity, a dark energy component with
negative pressure in the universe has been invoked as the most
feasible mechanism for the acceleration. In addition to the
cosmological constant (the $\Lambda$CDM model), many candidates of
dark energy have been taken into account. Examples include the
scalar field models with dynamical equation of state [e.g.,
quintessence (Ratra and Peebles 1988; Caldwell et al. 1998;
Choudhury and Padmanabhan 2005), phantom (Caldwell 2002; Wu and Yu
2006), quintom (Feng et al. 2005;  Guo et al. 2005;  Liang et al.
2009), k-essence (Armendariz-Picon et al. 2001; Chiba 2002), tachyon
(Padmanabhan 2002; Frolov et al. 2002)], the Chaplygin gas
(Kamenshchik et al. 2001) and the generalized Chaplygin gas model
(GCG, Bento et al. 2002; Zhu 2004), the holographic dark energy
(Cohen 1999; Li 2004), the agegraphic dark energy (Cai 2007; Wei \&
Cai 2008), the Ricci dark energy (Gao et al. 2009) and so on.

On the other hand, many alternatives to dark energy in which gravity
is modified have been proposed as a possible explanation for the
acceleration. Examples include the $f(R)$ theory in which the
Einstein-Hilbert action has been modified (Capozziello \& Fang 2002;
Vollick 2003; Carroll et al. 2004); the Cardassian expansion model
in which the Friedmann equation is modified by adding an extral
Cardassian term (Freese and Lewis 2002; Wang  et al. 2003; Zhu and
Fujimoto 2002, 2003); as well as the braneworld models, in which our
observable universe is considered as a brane embedded in a higher
dimensional bulk spacetime  and the leakage of gravity force
propagating into the bulk can lead to the current accelerated
expansion of the universe (Randall and Sundrum 1999).

In 2000, Dvali, Gabadadze and Porrati proposed a 5-dimensional brane
world model in which a self-accelerating branch is included (the
so-called DGP model, Dvali et al. 2000). The dynamics of gravity is
governed by a competition between a Ricci scalar term in the
4-dimensional brane and an Einstein-Hilbert action in the
5-dimensional bulk. The Friedmann equation of the DGP model is
modified as
\begin{equation}
H^2 = H_0^2 \left[ \Omega_\mathrm{K}(1+z)^2+\left(\sqrt{\2}+
        \sqrt{\2+\1 (1+z)^3}\right)^2 \right],
\end{equation}
where $H$ is the Hubble parameter as a function of redshift $z$,
$\1$ and $\Omega_\mathrm{K}$ represent the fractional contribution
of the matter and curvature, and $ \2 ={1}/{4r_{\rm c}^2H_0^2}$ is
the bulk-induced term respect to the crossover radius $r_{\rm c}$.
For scales below $r_{\rm c}$, the induced 4-dimensional Ricci scalar
dominates and the gravitational force is the usual $1/r^{2}$
behavior; whereas for distance scales larger than $r_{\rm c}$, the
gravitational force follows the 5-dimensional $1/r^{3}$ behavior.
The normalization condition can be given by
$\Omega_\mathrm{K}+\left(\sqrt{\2}+\sqrt{\2+\1}\right)^2 =1$; for a
spatially flat scenario, $\2=(1-\1)^2/4$.

The DGP model is a testable scenario with the same number parameters
as the standard $\Lambda$CDM model  and has been constrained from
many observational data, such as SNe Ia (Deffayet et al. 2002;
Avelino and Martins 2002; Zhu and Alcaniz 2005;  Maartens and
Majerotto 2006; Barger et al. 2007; Reboucas 2008), the angular size
of compact radio sources (Alcaniz 2002), the baryon mass fraction in
clusters of galaxies (CBF) from the x-ray gas observation (Zhu and
Alcaniz 2005; Alcaniz and Zhu 2005),  CMB (Lazkoz et al. 2006;
Rydbeck et al. 2007; He et al. 2007), the large scale structures
(Multam\"aki et al. 2003; Lue et al. 2004; Koyama and Maartens 2006;
Song et al. 2007) and the baryon acoustic oscillation (BAO) peak
(Guo et al. 2006), the observed Hubble parameter $H(z)$ data (Wan et
al. 2007), the gravitational lensing surveys (Jain et al. 2002; Zhu
and  Sereno 2008),  the age measurements of high-$z$ objects
(Alcaniz, Jain and Dev 2002) and the lookback time to galaxy
clusters (Pires, Zhu and Alcaniz 2006); as well as some different
combined data (Bento et al. 2006; Davis et al. 2007; Movahed et al.
2009; Xia 2009; Li et al. 2010). See Lue (2006) for review on the
DGP phenomenology.

Recently,  Gamma-ray bursts (GRBs) have been proposed as distance
indicators and regarded as a complementary cosmological probe to the
universe at high redshift (Schaefer 2003; Bloom et al. 2003; Dai et
al. 2004; Ghirlanda et al. 2004; Friedman and Bloom 2005;  Firmani
et al. 2005, 2006; Liang and Zhang 2005; Bertolami and Silva 2006;
Ghirlanda et al. 2006; Schaefer 2007; Wright 2007; Wang et al. 2007;
Amati et al. 2008; Basilakos and Perivolaropoulos 2008; Mosquera
Cuesta et al. 2008a, 2008b; Qi et al. 2008a, 2008b). For constraints
on the DGP model from GRBs with their associated joint observations,
see some recent works, e.g., Wang et al. (2009a); Wei (2010a); Xu
and Wang (2010). However, the empirical luminosity relations of GRBs
have usually been  calibrated by assuming a certain cosmological
model with particular model parameters, due to the lack of the
low-redshift sample. Therefore the calibration are always
cosmology-dependent and the so-called circularity problem occurs in
GRB cosmology. The circularity problem cannot be avoided completely
by means of statistical approaches (Schaefer 2003; Li et al. 2008;
Wang 2008; Samushia and Ratra 2010; Xu 2010), because an input
cosmological model is still required. Liang et al. (2008) presented
a new method to calibrate GRB luminosity relations in a completely
cosmology-independent way: GRB sample in the redshift range of SNe
Ia are enough to calibrate GRB relations and their luminosity
distances can be obtained directly from SNe Ia by the interpolation
method or  by other similar approach (Liang and Zhang 2008; Kodama
et al. 2008; Cardone et al. 2009; Gao et al. 2010; Capozziello and
Izzo 2010). Following the cosmology-independent GRB calibration
method, the derived GRB data at high redshift can be used to
constrain cosmological models by using the standard Hubble diagram
method (Capozziello and Izzo 2008; Izzo et al. 2009; Wei and Zhang
2009; Wei 2009; Qi et al. 2009; Wang et al. 2009a, 2009b; Wang and
Liang 2010; Liang, Wu and Zhang 2010; Liang, Wu and Zhu 2010; Wei
2010a, 2010b; Freitasa et al. 2010; Liang, Xu and Zhu 2010;
Demianski et al. 2010).

Very recently, Liang, Wu and Zhu (2010) calibrated GRB data at high
redshift directly from the Union2 compilation of 557 SNe Ia data set
(Amanullah et al. 2010); and constrained the Cardassian model
(Liang, Wu and Zhu 2010) and the generalized Chaplygin gas (GCG)
model (Liang, Xu and Zhu 2010) by combining the updated GRB data
with the joint observations, such as the Union2 set of SNe Ia, the
CMB observation from the seven-year data of Wilkinson Microwave
Anisotropy Probe (WMAP7; Komatsu et al. 2010) result and the BAO
observation from the spectroscopic Sloan Digital Sky Survey (SDSS)
galaxy sample (Eisenstein et al. 2005).

In this paper, we investigate observational constraints on the DGP
model including the updated the distance moduli of the GRBs at high
redshift obtained directly from the Union2 set. We combine the GRB
data with the joint observations such as the Union2 set, the CMB
observation from the WMAP7 result; the BAO observation from the
spectroscopic SDSS galaxy sample (Eisenstein et al. 2005); the
baryon mass fraction in clusters of galaxies from the x-ray gas
observation (Allen et al. 2004); and the observed Hubble parameter
data ($H(z)$; Simon et al. 2005; Gazta$\tilde{n}$aga et al. 2009).
We also  obtain the transition redshift $z_\mathrm{T}$ of the DGP
model. We find that the combination of these recent data sets
tighter constraints on the DGP model, which favors a flat universe.
The paper is organized as follows. 
In section 2, we introduce the analysis for the observational data
including the updated cosmology-independent GRBs, as well as the
Union2 SNe Ia set, the CMB observations from the WMAP7 result, in
addition to the BAO, CBF and $H(z)$ data. In section 3, we present
results which put constraints on the DGP model from the joint
observations. Conclusions and discussions are given in section 4.

%
%

\section{Observational Data Analysis}
The recent Union2 compilation consists of a 557 SNe Ia data set
(Amanullah et al. 2010). In this paper,  we use the updated the
distance moduli of the 42 GRBs at $z>1.4$ (Liang, Wu and Zhu 2010),
which are obtained by the five luminosity relations (Schaefer 2007)
calibrated with the sample at $z\le1.4$ by using the linear
interpolation method from the Union2 set. For more details about the
calculation, see (Liang et al. 2008; Liang, Wu and Zhang 2010).
Constraints from SNe Ia and GRB data can be obtained by fitting the
distance moduli $\mu(z)$. A distance modulus can be calculated as
\begin{eqnarray}\label{mu}
\mu=5\log \frac{d_L}{\textrm{Mpc}} + 25=5\log_{10}D_L-\mu_0,
\end{eqnarray}
where $\mu_0=5\log_{10}[H_0/(100{\rm km/s/Mpc})]+42.38$, and the
luminosity distance $D_L$ is calculated by
\begin{eqnarray}
D_L\equiv
H_0d_L=(1+z)\Omega_\mathrm{k}^{-1/2}\textrm{sinn}\bigg[\Omega_\mathrm{k}^{1/2}\int_0^z\frac{dz'}{E(z')}\bigg],
\end{eqnarray}
where $\rm{sinn}$$(x)$ is $\rm sinh$ for $\Omega _{\rm k}>0$, $\rm
sin$ for $\Omega _{\rm k}<0$, and $x$ for $\Omega _{\rm k}=0$. The
$\chi^2$ value of the observed distance moduli can be calculated by
\begin{eqnarray}
\chi^2_{\mu}=\sum_{i=1}^{N}\frac{[\mu_{\mathrm{obs}}(z_i)-\mu(z_i)]^2}
{\sigma_{\mu,i}^2},
\end{eqnarray}
where $\mu _{\mathrm obs}(z_i)$ are the observed distance modulus
for the SNe Ia and/or GRBs at redshift $z_i$ with its error
$\sigma_{\mu_{\mathrm i}}$; $\mu(z_i)$ are the theoretical value of
distance modulus from cosmological models. Following an effective
approach (Nesseris and Perivolaropoulos 2005), we marginalize the
nuisance parameter $\mu_0$ by minimizing
\begin{eqnarray}
{\hat\chi}^2_{\mu}=C- {B^2}/{A},
\end{eqnarray}
where $A=\sum{1}/{\sigma_{\mu_{i}}^2}$,
$B=\sum{[\mu_{\mathrm{obs}}(z_i)-5\log_{10}D_L]}/{\sigma_{\mu_{i}}^2}$,
and
$C=\sum{[\mu_{\mathrm{obs}}(z_i)-5\log_{10}D_L]^2}/{\sigma_{\mu_{i}}^2}$.

For the CMB observation from the WMAP7 result (Komatsu et al. 2010),
the shift parameter is constrained to be $R=1.725\pm0.018$, which
can be expressed as (Bond et al. 1997)
\begin{equation}
R=\Omega_{\mathrm{M0}}^{1/2}\Omega_\mathrm{k}^{-1/2}\textrm{sinn}\bigg[\Omega_\mathrm{k}^{1/2}\int_0^{z_{\mathrm{rec}}}\frac{dz}{E(z)}\bigg],
\end{equation}
where $z_{\rm rec}$ is the redshift of recombination which is given
by (Hu and Sugiyama 1996)
\begin{equation}
z_{\mathrm{rec}}=1048[1+0.00124(\Omega_bh^2)^{-0.738}(1+g_{1}(\Omega_{\mathrm{M0}}h^2)^{g_2})],
\end{equation}
where
$g_1=0.0783(\Omega_bh^2)^{-0.238}(1+39.5(\Omega_bh^2)^{-0.763})^{-1}$
and $g_2=0.560(1+21.1(\Omega_bh^2)^{1.81})^{-1}$. From the WMAP7
result (Komatsu et al. 2010), $z_\mathrm{rec}=1091.3$. The $\chi^2$
value of the shift parameter can be calculated by
\begin{equation}
\chi^2_{\mathrm{CMB}}=\frac{(R-1.725)^2}{0.018^2}.
\end{equation}

For the BAO observation from the SDSS spectroscopic sample of
luminous red galaxy, the distance parameter is measured to be $A =
0.469(n_s/0.98)^{-0.35}\pm0.017$ (Eisenstein et al. 2005), with the
scalar spectral index $n_s=0.963$ from the WMAP7 result (Komatsu et
al. 2010). The distance parameter can be expressed as
\begin{equation}
A={\Omega_{\mathrm{M0}}^{1/2}}{z_\mathrm{BAO}^{-2/3}E(z_\mathrm{BAO})^{-1/3}}\Omega_\mathrm{k}^{-1/2}\textrm{sinn}\bigg[\Omega_\mathrm{k}^{1/2}\int_0^{z_\mathrm{BAO}}\frac{dz}{E(z)}\bigg]^{2/3}
\end{equation}
where $z_{\mathrm{BAO}}=0.35$. The $\chi^2$ value of the distance
parameter can be calculated by
\begin{equation}
\chi^2_{\mathrm{BAO}}=\frac{(A-0.467)^2}{0.017^2}.
\end{equation}

The  baryon  mass fraction  in clusters of galaxies from the x-ray
gas ($f_{\mathrm{gas}}$) observation can be used to constrain
cosmological parameters. On the assumption that the gas mass
fraction in clusters is a constant and thus independent of redshift,
Allen et al. (2004) obtained 26 observational $f_{\mathrm{gas}}$
data. The baryon gas mass fraction $f_{\mathrm{gas}}$ can be
presented as
\begin{equation}
f_{\mathrm{gas}}(z)=\lambda\bigg[\frac{d_A^{\mathrm{SCDM}}(z)}{d_A(z)}\bigg]^{2/3},
\end{equation}
where  $d_A\equiv d_L/(1+z)^2$ is the theoretical value of the
angular diameter distance from cosmological models,
$d_A^{\mathrm{SCDM}}$ is the angular diameter distance corresponding
to the standard cold dark matter model (SCDM, $\Omega_{\mathrm{M0}}
= 1$ for a flat universe), and
$\lambda=[b\Omega_\mathrm{b}(2h)^{3/2}]/[(1+a)\Omega_{\mathrm{M0}}]$,
$a=0.19\sqrt{h}$, $b$ is a bias factor motivated by gas dynamical
simulations. The $\chi^2$ value of cluster's baryon gas mass
fraction (CBF) is
\begin{eqnarray}
\chi^2_{\mathrm{CBF}}=\sum_{i=1}^{N=26}\frac{[f_{\mathrm{gas}}^{\mathrm{obs}}(z_i)-f_{\mathrm{gas}}(z_i)]^2}
{\sigma_{f_{\mathrm{gas}},i}^2}.
\end{eqnarray}
The parameter $\lambda$ can be  treated as a nuisance parameter by
minimizing (Nesseris  and Perivolaropoulos 2007)
\begin{eqnarray}
{\hat\chi}^2_{\mathrm{CBF}}=C- {B^2}/{A},
\end{eqnarray}
where
$A=\sum[{\tilde{f}_{\mathrm{gas},i}}/\sigma_{f_{\textrm{gas}},i}]^2$,
$B=\sum{[\tilde{f}_{\mathrm{gas},i}f_{\mathrm{gas},i}]}/{\sigma_{f_{\mathrm{gas}},i}^2}$,
$C=\sum{[f_{\mathrm{gas},i}}/\sigma_{f_{\mathrm{gas}},i}]^2$, and
${\tilde{f}_{\mathrm{gas},i}}=[{d_A^{\mathrm{SCDM}}(z)}/{d_A(z)}]^{2/3}$.

The Hubble parameter $H(z)$ can be derived by
\begin{equation}
H(z)=-\frac{1}{1+z}\frac{dz}{dt}.
\end{equation}
From the Gemini Deep Deep Survey (GDDS; Abraham et al. 2004)
observations of differential ages of passively evolving galaxies and
other archival data (Nolan et al. 2003; Treu et al. 2001, 2002;
Spinrad et al. 1997; Dunlop et al. 1996), Simon et al. (2005) have
obtained  the $H(z)$ data at nine different redshifts ($0.09\leq
z\leq1.75$).  Recently, $H(z)=83.2\pm 2.1 {\rm km/s/Mpc}$ at
$z=0.24$, and $H(z)=90.3\pm 2.5 {\rm km/s/Mpc}$ at $z=0.43$ have
been obtained by using the BAO peak position as a standard ruler in
the radial direction (Gazta$\tilde{n}$aga et al. 2009). The $\chi^2$
value of the 11 $H(z)$ data is
\begin{eqnarray}
\chi^2_{H}=\sum_{i=1}^{N=11}\frac{[H_{\mathrm{obs}}(z_i)-H(z_i)]^2}
{\sigma_{H,i}^2}.
\end{eqnarray}
The nuisance parameter $H_0$ is also marginalized following the
procedure used in calculating $\hat\chi^2_{\mu}$.

\section{CONSTRAINTS FROM COMBINING GRBs, SNe Ia, CMB, and BAO }
In order to combine GRB data with the SNe Ia data to constrain
cosmological models, we follow a simple way that avoids any
correlation between the SNe Ia data and the GRB data (Liang, Wu and
Zhang 2010): The 40 SNe points used in the interpolation procedure
to calibrate GRBs are excluded  from the Union2 SNe Ia sample used
to calculate the joint constraints. Since the reduced 517 SNe Ia, 42
GRBs, CMB, BAO, as well as CBF and $H(z)$ are all effectively
independent, we can combine the results by simply multiplying the
likelihood functions. The best fit values for model parameters from
the distance moduli of GRBs at high redshift obtained directly from
the Union2 set, and SNe Ia,  as well as the other joint observations
(CMB+BAO+CBF+$H(z)$) can be determined by minimizing
\begin{equation}
\chi^2=\hat\chi^2_{\mu,\{\rm{42GRBs+517SNe}\}}+\chi^2_{\mathrm{CMB}}+\chi^2_{\mathrm{BAO}}+{\hat\chi}^2_{\mathrm{CBF}}+\hat{\chi}^2_{H}\;.
\end{equation}
In order to show the contribution of GRBs to the joint cosmological
constraints, we also consider  the $\chi^2$ value from the joint
data (SNe + CMB + BAO + CBF + $H(z)$) without GRBs:
$\chi^2_{S}=\hat\chi^2_{\mu,\{\rm{557SNe}\}}+\chi^2_{\mathrm{CMB}}+\chi^2_{\mathrm{BAO}}+{\hat\chi}^2_{\mathrm{CBF}}+\hat{\chi}^2_{H}$;
and  the joint constraints with GRBs + CMB + BAO + CBF + $H(z)$
without the SNe Ia contribution is:
$\chi^2_{G}=\hat\chi^2_{\mu,\{\rm{42GRBs}\}}+\chi^2_{\mathrm{CMB}}+\chi^2_{\mathrm{BAO}}+{\hat\chi}^2_{\mathrm{CBF}}+\hat{\chi}^2_{H}$.

The joint confidence regions in
$\{\Omega_{\mathrm{M0}}$-$\Omega_\mathrm{r_{\rm c}}\}$ plane with
the combined observational data for the DGP model are showed in
figure 1.  For comparison, fitting results from the joint data
without GRBs are also given in figure 1. We present the best-fit
value of $\{ \1,\2 \}$ with 1-$\sigma$ uncertainties and the
corresponding $\Omega_{\mathrm{K}}$, as well as $\chi_{\rm min}^2$,
$\chi_{\rm min}^2/\textrm{dof}$ for the DGP model in Table 1. We
also investigate the deceleration parameter for the DGP model. The
deceleration parameter $q(z)$  can be calculated by
$q=-1+(1+z)E(z)^{-1}{dE(z)}/{dz}$, where  $E(z)=H/H_0$. And we could
derive the  transition redshift at which the  universe of the DGP
model switches from deceleration to acceleration (Zhu and Alcaniz
2005; Guo et al. 2006)
\begin{equation}
z_\mathrm{T}= -1+2\big(\frac{\2}{\1}\big)^{1/3}
\end{equation}
The best-fit values of  $z_\mathrm{T}$ of the DGP model are also
summarized in Table 1.

With SNe Ia + GRBs + CMB + BAO + CBF + $H(z)$, the best-fit values
at 1-$\sigma$ confidence level are $\{ \1, \2 \}
=\{0.235_{-0.074}^{+0.125}, 0.138_{-0.036}^{+0.031}\}$; with the
corresponding $\Omega_{\mathrm{K}}\simeq 0$, which is near the line
of a flat universe ($(1-\1)^2-4\2=0$); with SNe Ia + CMB + BAO + CBF
+ $H(z)$, the best-fit values are $\{ \1, \2 \}
=\{0.217_{-0.073}^{+0.126}, 0.144_{-0.035}^{+0.032}\}$; while with
GRBs + CMB + BAO + CBF + $H(z)$, the best-fit values are $\{ \1, \2
\} =\{0.285_{-0.066}^{+0.252}, 0.122_{-0.062}^{+0.044}\}$. These
results lead to more stringent constraints than previous results for
constraint on DGP model with GRBs and/or other combined observations
(Wang et al. 2009a; Wei 2010a; Bento et al. 2006; Davis et al. 2007;
Reboucas 2008; Li et al. 2010). We also obtain the transition
redshift $z_\mathrm{T}=0.67_{-0.04}^{+0.03}$ ($1 \sigma$) with the
joint data including GRBs, which is more stringent and later the
former result ($z_\mathrm{T}=0.86_{-0.08}^{+0.07}$) in Guo et al
(2006).

 \begin{table}[tbhp]
 \begin{center}
 \begin{tabular}{|c|c|c|c|} \hline\hline
 & \multicolumn{3}{c|}{The DGP Model}  \\
 \cline{2-4}                         &     SNe+GRBs+Others                 &     SNe+Others                      &     GRBs+Others         \\ \hline
 $\Omega_{\mathrm{M0}}$          \ \ & \ \ $0.235_{-0.074}^{+0.125}$\ \  & \ \ $0.217_{-0.073}^{+0.126}$\ \   & \ \ $0.285_{-0.066}^{+0.252}$\ \ \\
 $\Omega_\mathrm{r_{\rm c}}$     \ \ & \ \ $0.138_{-0.036}^{+0.031}$\ \  & \ \ $0.144_{-0.035}^{+0.032}$\ \   & \ \ $0.122_{-0.062}^{+0.044}$\ \ \\
 $\Omega_\mathrm{k}$             \ \ & \ \ $0.033$\ \                    & \ \ $0.037$\ \                     & \ \ $0.024$\ \ \\ \hline
 $\chi_{\rm min}^2$              \ \ & \ \ $595.95$\ \                   & \ \ $606.37$\ \                   & \ \ $77.93$\ \      \\
 $\chi_{\rm min}^2/\textrm{dof}$ \ \ & \ \ $1.07$\ \                     & \ \ $1.09$\ \                     & \ \ $0.99$\ \ \\  \hline\hline
 $z_\mathrm{T}$                  \ \ & \ \ $0.67_{-0.04}^{+0.03}$\ \     & \ \ $0.74_{-0.07}^{+0.05}$\ \     & \ \ $0.51_{-0.16}^{+0.14}$\ \ \\
 \hline\hline
 \end{tabular}
 \end{center}
 \caption{\label{tab2} The best-fit value of the DGP model  parameters   $\{\Omega_{\mathrm{M0}}$-$\Omega_\mathrm{r_{\rm
c}}\}$ and  $\Omega_\mathrm{k}$  with 1-$\sigma$ uncertainties,
$\chi_{\rm min}^2$, $\chi_{\rm min}^2/\textrm{dof}$, as well as
$z_\mathrm{T}$ with SNe+GRBs+Others (CMB+BAO+ CBF+$H(z))$,
SNe+Others, and GRBs+Others, respectively.}
 \end{table}
\begin{figure}
\centering
\includegraphics[angle=0,width=0.6\textwidth]{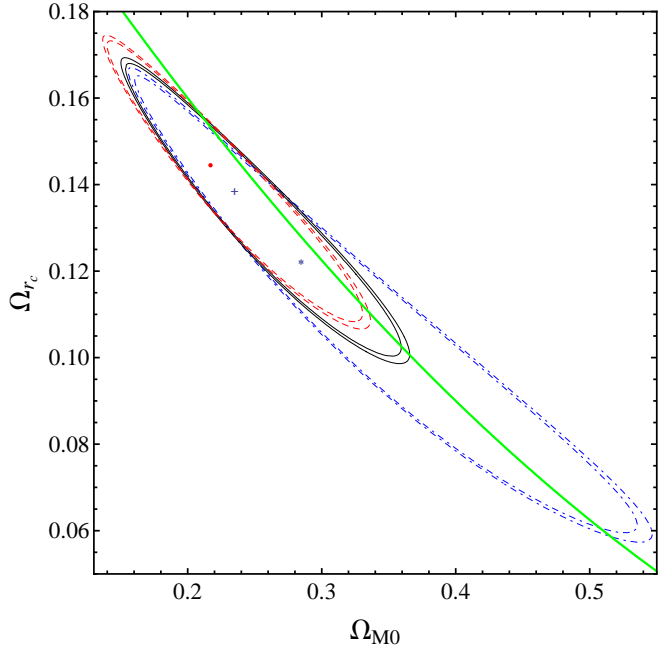}
\caption{The joint confidence regions in $\{ \1$-$\2 \}$ plane for
the DGP model. The contours correspond to 1-$\sigma$ and 2-$\sigma$
confidence regions. The black solid lines, red dashed lines, and the
blue dash-dotted lines represent the results of
SNe+GRBs+Others(CMB+BAO+ CBF+$H(z))$, SNe+CMB+Others, and
GRBs+Others, respectively. The black plus, red point, and blue star
correspond the best-fit values of SNe+GRBs+Others, SNe+Others and
GRBs+Others, respectively. The green line represents a flat universe
which can be given by $(1-\1)^2-4\2=0$. }
\end{figure}

From comparing to the joint constraints with GRBs and without GRBs,
we can see that the contribution of GRBs to the joint cosmological
constraints of the DGP model is a shift between the best fit values
near the line which represents a flat universe, towards a higher
matter density Universe ($\Delta\Omega_{\rm M0}>0$). This situation
has been also noted by Liang, Wu and Zhang (2010), and Liang, Wu and
Zhu (2010), who comparing to the joint constraints with GRBs and
without GRBs using the $\Lambda$CDM model, $w$CDM model, and
Cardassian model. Also, a shift towards a later transition redshift
can be found by comparing to the joint constraints of the DGP model
with and without GRBs. It is  shown that  GRBs can give strong
constraints on the DGP model when combined with CMB and BAO
observations without SNe Ia, which has been also noted by Liang, Wu
and Zhu (2010); Liang, Xu and Zhu (2010).

\section{Conclusion and discussion}
In this paper, we investigate observational constraints on the DGP
model including the cosmology-independent GRBs obtained directly from SNe Ia. 
Combining the GRBs at high redshift with the Union2 set, the WMAP7
result, the BAO observation, the clusters' baryon mass fraction, and
the observed Hubble parameter data,  we obtain $\{ \1,\2 \}
=\{0.235_{-0.074}^{+0.125},0.138_{-0.036}^{+0.031}\}$, with the
corresponding $\Omega_{\mathrm{K}}=0.033$, which favors a flat
universe. We also obtain the transition redshift of the DGP model
$z_\mathrm{T}=0.67_{-0.04}^{+0.03}$. These results breaks the
degeneracies between the model parameters and leads to more
stringent constraints than the previous results for constraint on
DGP model  with GRBs and/or other combined observations. It is shown
that GRBs can give strong constraints on the DGP model when combined
with CMB and BAO observations. We conclude that GRBs could be used
as an optional choice to set tighter constraints at high redshift on
cosmological models.

Zhu and Alcaniz (2005) tested the DGP model with the baryon mass
fractions in clusters of galaxies and the SNe Ia data to find that
$\{ \1\,\2 \}= \{0.29^{+0.04}_{-0.02}, 0.21^{+0.08}_{-0.08}\}$, and
$\Omega_k=-0.36^{+0.31}_{-0.35}$ at 99.73\% confidence level. Guo et
al. (2006) also obtained a spatially closed DGP universe with
$\Omega_k=-0.350^{+0.080}_{-0.083}$ by using SNe Ia + BAO data. Zhu
and Sereno (2008) used gravitational lensing statistics to find that
the likelihood peaks at $\{ \1,\2 \} \simeq \{0.29, 0.12\}$, just
slightly in the region of open models. These results seem to be in
contradiction with the most recent WMAP results indicating a flat
universe. However, constraints on the DGP model of the joint data
including GRBs in this work  are consistent with those obtained by
Bento et al. (2006) using SNe Ia + CMB + BAO, and by Reboucas (2008)
using SNe Ia + CMB; which favor a flat universe.

\begin{acknowledgements}
We thank Yun Chen, He Gao, Shuo Cao, Hao Wang, Yan Dai, Chunhua Mao,
Fang Huang, Yu Pan, Jing Ming, Kai Liao  and Dr. Yi Zhang for
discussions. This work was supported by the National Science
Foundation of China under the Distinguished Young Scholar Grant
10825313, the Key Project Grants 10533010, and by the Ministry of
Science and Technology national basic science Program (Project 973)
under grant No. 2007CB815401.
\end{acknowledgements}


\end{document}